\documentclass[12pt]{article}
\usepackage{epsfig}
\usepackage{graphicx}
\newcommand{\be}{\begin{equation}}
\newcommand{\ee}{\end{equation}}
\newcommand{\ba}{\begin{eqnarray}}
\newcommand{\ea}{\end{eqnarray}}
\newcommand{\cao}{{\cal O}}
\newcommand{\tao}{\tilde{{\cal O}}}

\newcommand{\vp}{{\vec p}}
\newcommand{\noi}{\noindent}
\newcommand{\bpartial}{{\bar \partial}}
\begin{document}
\renewcommand{\baselinestretch}{1.3}
\small\normalsize
\renewcommand{\theequation}{\arabic{section}.\arabic{equation}}
\renewcommand{\thesection}{\arabic{section}.}
\language0

\thispagestyle{empty}

\begin{flushleft}
February 1998 \hfill HUB-EP-98/10 \\
                \hfill JINR E2-98-27
\end{flushleft}

\vspace*{1.0cm}

\begin{center}

{\Large \bf Gluon propagator and zero-momentum modes\\
in SU(2) lattice gauge theory}

\vspace*{0.8cm}

{\bf G.~Damm $^a$ , W.~Kerler $^{a,b}$ , V.K.~Mitrjushkin $^c$}

\vspace*{0.3cm}

\end{center}

{\sl \noindent
 \hspace*{6mm} $^a$ Fachbereich Physik, Universit\"at Marburg,
		       D-35032 Marburg, Germany \\
 \hspace*{6mm} $^b$ Institut f\"ur Physik, Humboldt-Universit\"at,
		       D-10115 Berlin, Germany\\
 \hspace*{6mm} $^c$ Joint Institute for Nuclear Research, Dubna, Russia}

\vspace*{1.5cm}

\begin{abstract}
We investigate propagators in Lorentz (or Landau)
gauge by Monte Carlo simulations. In order to be able to compare with
perturbative calculations we use large $\beta$ values. There the
breaking of the $Z(2)$ symmetry turns out to be important for all of the
four lattice directions. Therefore we make sure that the analysis is
performed in the correct state. We discus implications of the gauge
fixing mechanism and point out the form of the weak-coupling behavior to
be expected in the presence of zero-momentum modes. Our numerical result
is that the gluon propagator in the weak-coupling limit is strongly
affected by zero-momentum modes. This is corroborated in detail by our
quantitative comparison with analytical calculations.  \end{abstract}

\newpage
\renewcommand{\baselinestretch}{1.37}
\small\normalsize

\section{Introduction} \setcounter{equation}{0}

Fundamental results on QCD have been obtained in continuum perturbation
theory as well as nonperturbatively on the lattice. It appears important
to connect the knowledge from these two approaches more closely.
Progress within this respect is expected from the calculations of
typical quantities of perturbation theory by Monte Carlo simulations in
lattice gauge theory.
The gluon propagator is a fundamental quantity for such an investigation.

In the seminal paper by Mandula and Ogilvie \cite{mo87} the
zero--momentum gluon propagator
$~\Gamma(\tau)=\Gamma(\tau;\vp={\vec 0})~$ has been
numerically calculated in the Lorentz (or Landau) gauge.
The analysis of the $~\tau$--dependence of this propagator drove the
authors of \cite{mo87} to the conclusion that $~\Gamma(\tau)~$ is
consistent with the propagation of a massive gluon. Therefore,
the effect of dynamical mass generation was declared to be observed.
Since that there was quite a number of papers devoted
to the calculation and analysis of gauge--variant zero--momentum
propagators at zero and nonzero temperatures following the same line
as paper \cite{mo87} (see, e.g.~[2--8]) 
However, the impact of zero--momentum modes on the propagators
has not been analyzed.

Recently it has been shown by one of us that zero-momentum modes
may strongly affect some gauge-dependent correlators \cite{m97}.
In some cases, e.g., in the Coulomb phase in pure $~U(1)~$ gauge 
theory, zero--momentum modes can mimic the effective 
masses defined in a standard way (see eq.~(\ref{m})).
This is apparent from the data of \cite{mo2}.
Because so far nothing is known about the magnitude of this
effect in nonabelian gauge theory, it appears highly desirable 
to perform simulations which clarify this issue.
In order to be able to compare quantitatively with perturbative
calculations and to draw firm conclusions it is reasonable to begin
this study in the truly perturbative region, i.e. at large values 
of $\beta$.

In the present paper we perform Monte Carlo simulations in pure $SU(2)$
gauge theory in the four--dimensional volume $~V_4=L_1L_2L_3L_4~$ 
with periodic boundary conditions.  We use the Wilson action 
\be
S=\beta\sum_P \Bigl(1-\frac{1}{2}\mbox{Tr}\, U_P \Bigr)~,
\ee
\noi and the Lorentz gauge as described in Section 2. 
We discuss the working of the gauge fixing mechanism in detail
and present analytical forms of the propagators appropriate at
weak-coupling in the presence of zero-momentum modes. We demonstrate the
effects of the broken $~Z(2)~$ symmetry and show that in the appropriate
$~Z(2)~$ state there is agreement with perturbative calculations. In
particular, we find that there is a sizable effect caused by
zero-momentum modes which can be understood quantitatively. 

\section{Gauge fixing procedure} \setcounter{equation}{0}

In $SU(2)$ lattice gauge theory the Lorentz gauge is fixed by maximizing
\be
F=\frac{1}{4V_4} \sum_{\mu,x} \mbox{Tr }U_{\mu x}
\label{F}
\ee

\noi by appropriate gauge transformations
$U_{\mu x} \rightarrow V_{x+e_{\mu}} U_{\mu x} V_{x}^{\dag}$ ,
which implies the local condition
\be
\sum_{\mu} \bpartial_{\mu}U_{\mu x}=0
\ee
(where $\bpartial_{\mu}f(x)=f(x)-f(x-e_{\mu})$). The maximized functional
(\ref{F}) may be cast into the form
\be
F_{\mbox{\scriptsize max}}= \frac{1}{8V_4} \sum_x \mbox{Tr }(M_xV^{\dag}_x)
\label{Fm}
\ee
where
\be
M_{x}=\sum_{\mu}
(V_{x+e_{\mu}}U_{\mu x}+V_{x-e_{\mu}}U_{\mu,x-e_{\mu}}^{\dag}) \quad .
\label{sol}
\ee
 From the fact that (\ref{Fm}) is maximal 
it follows that the maximizing gauge transformations are given by
\be
V_{x}=M_{x}/\|M_{x}\|
\label{Vx}
\ee
where $\|M_{x}\|^2=\frac{1}{2}\mbox{Tr}(M_{x}M_{x}^{\dag})$. Obviously
(\ref{Vx}), 
is a system of equations which determines all $V$'s given the $U$'s
of the configuration. (It should not to be mixed up with the relation without
$V$'s on the r.h.s. and with the $U$'s being the ones of the particular
iteration step, which is used iteratively site by site in numerical gauge
fixing).

The gauge--fixing procedure is based on the equality
      $~\langle P \rangle_f = \langle P^V \rangle~$
of the expectation with gauge fixing for the observable $P$ to that
without gauge fixing for the transformed observable $P^V$. We note that
this actually amounts to switching to the gauge--invariant observable
$P_{\mbox{\scriptsize eff}}$ which is related to the gauge-variant one
$P$ in the gauge under consideration, i.e.~that $P_{\mbox{\scriptsize
eff}}=P^V$ . The respective general relation
  $~ \langle P \rangle_f = \langle P_{\mbox{\scriptsize eff}} \rangle~$
with $P_{\mbox{\scriptsize eff}}=\int_V {\cal F} P / \int_V {\cal F}$ ,
where ${\cal F}$ is a general gauge fixing function, has been pointed
out some time ago \cite{k81}.

It should be realized that the mechanism leading from a gauge-variant
observable $P$ to an effectively gauge-invariant one 
$P_{\mbox{\scriptsize eff}}$ is simple. The gauge-fixing function
provides strings of gauge fields which can combine with the given 
gauge-variant operators to gauge-invariant objects (and, of course,
only such objects contribute to the integrals). For example, given
a gauge field at a certain link, a string may be provided such that 
one gets a loop which contains this link. In practice there can be
many suitable strings such that one arrives at a sum of gauge-invariant
terms.

In the present case the gauge-invariant observables can be constructed.
To see this one has to realize that after maximization the original $U$
variables due to the gauge transformation are decorated by $V$ factors.
On the other hand, these $V$ factors are given here by (\ref{Vx}), 
which therefore can be used to eliminate the $V$'s successively from $P^V$. 
In detail, one replaces a $V$ by (\ref{Vx}) which introduces next $V$'s
from the r.h.s~of (\ref{Vx}), then replaces these $V$'s again by (\ref{Vx}), 
and so on. This procedure terminates when all $V$'s are eliminated and 
one has arrived at an expression which contains only closed loops of 
gauge fields $U$.

In the gauge-invariant observable thus obtained there are loops which
include the gauge-variant parts of the gauge-variant observable $P$
according to the mechanism explained above. Furthermore, noting that
the factors $\|M_x\|$ in (\ref{Vx}) also contain $V$'s, it becomes obvious
that the terms of the effectively invariant observable are multiplied by 
weights which stem from these $\|M_x\|$ factors and which only depend on 
gauge-field loops (the locations of which are not restricted). 

This construction allows insight into properties to be expected. First
of all it is obvious that very large loops and, in particular, also
loops of Polyakov type occur. Thus one can expect that the effective
gauge-invariant observables resulting from gauge-variant ones become
sensitive to the breaking of the $~Z(2)~$ symmetry. The
gauge-variant propagators investigated here indeed turn out to be
strongly affected by the occurring $~Z(2)~$ states as will be discussed
in Section 4.

In the case of the propagators defined in Section 3 $P$ contains 
gauge-variant terms of forms $\mbox{Tr}(U_{\mu x}U_{\nu y})$, 
$\mbox{Tr}(U_{\mu x}U_{\nu y}^{\dag})$ and their hermitian conjugates. 
By the maximizing gauge transformation then, for example, the term  
$\mbox{Tr}(U_{\mu x}U_{\nu y})$ gets the form 
$\mbox{Tr}(V_{x+e_{\mu}} U_{\mu x} V_{x}^{\dag} 
	   V_{y+e_{\mu}} U_{\mu y} V_{y}^{\dag})$. 
The elimination procedure of $V$'s described above forms loops of gauge fields
incorporating the $U$ factors of these terms. In the nonabelian case because of 
the product form of these terms one necessarily has to connect $x$ and $y$ 
in order to eliminate all $V$'s. In other words, one has to form loops
connecting $x$ and $y$ while separate loops through $x$ and $y$ are not 
possible. This has the consequence that one actually does not have 
correlations between separate operators, which implies that the usual
transfer-matrix considerations of spectral properties do not apply.
%
%

We have performed Monte Carlo simulations on lattices of sizes
$4^3\times 8$ , $8^3\times 16$ and $16^3\times 32$. After studies at
various $\beta$ the investigations to be described in Section 5 have
been at large $\beta$ (mainly at $\beta=10$, some at $\beta=99$).
Similarly, after studies of transition rates between $~Z(2)~$ symmetry
states with heat bath as well as Metropolis algorithms, for the main
simulations heat bath update has been used. Measurements have been
usually separated by 100 sweeps, on the small lattice also by 10 sweeps.
Gauge fixing has been done iteratively site by site using the local
solution (\ref{sol}) (with no $V$'s on the r.h.s. and with the $U$'s
being the ones of the particular iteration step) and applying stochastic
overrelaxation \cite{fg89}.  This iterative procedure has been stopped
if $\frac{1}{4V}\sum_x \|\sum_{\mu} \partial^-_{\mu}\cao_{\mu}(x)\|^2 <
10^{-8}$.

Statistical errors of observables have been determined by the jackknife
method.  To control the quality of gauge fixing 10 gauge copies have
been generated by random gauge transformations for the configurations
used in the measurements and the data separately analyzed for the best
(largest $F$), the first and the worst (smallest $F$) copy.  The
differences have been found to be small as compared to the statistical
errors of the observables of interest, which agrees with similar
observations in other investigations \cite{hkr95,hkr97,c97}.  Furthermore, it
has also been checked that $\Gamma_4(\tau)$ is constant within errors as
it should be.

\section{Forms of propagators} \setcounter{equation}{0}

Let us define traceless fields $~\cao_{\mu}(x)~$ 

\be
\cao_{\mu}(x)=\frac{1}{2i}(U_{\mu x}-U^{\dag}_{\mu x})~,
\qquad U_{x\mu} = e^{igA_{x\mu}}~,
\label{O}
\ee

\noi and consider propagators

\be
\Gamma_{\mu}(\vec{p},\tau)=\frac{1}{L_4}\sum_t \mbox{Tr}
\Bigl\langle\tao_{\mu}(\vec{p},t+\tau) \tao_{\mu}(-\vec{p},t) 
\Bigr\rangle
\label{prop}
\ee

\noi where (with $V_3=L_1 L_2 L_3$)

\be
\tao_{\mu}(\vec{p},\tau)=\frac{1}{V_3}\sum_{\vec{x}}e^{i\vec{p}\cdot\vec{x}}
			 \cao_{\mu}(\vec{x},\tau) \quad .
\ee
We choose $\vec{p}=(0,0,p_3)$ (where $p_3=\frac{2\pi}{L_3}\rho$ with
integer $\rho$) so that the transverse components are
$\Gamma_1$ and $\Gamma_2$ (in the following we use $\Gamma_T=\frac{1}{2}
(\Gamma_1+\Gamma_2)$ and analogous expressions for related quantities).

Perturbative results serve usually as a reference form in the numerical
(nonperturbative) study.  However, a perturbative approach in nonabelian
gauge theory enclosed in a periodic box faces a problem
\cite{baaq,lues}. This is the zero mode problem which makes a normal
perturbative expansion invalid because of the divergent gaussian
integrals. These zero modes (i.e. zero--eigenvalue eigenstates of the
quadratic form) are associated with zero momentum, so this problem is
usually referred to as zero--momentum mode problem.

Ignoring the divergency of the gaussian integrals (which is usually the
case) one arrives at the following expression for the free gluon
propagator:

\be
\Gamma_{\mu}(\vec{p},\tau) = \frac{3g^2}{2V_4}\sum_{p_4}
\frac{e^{-ip_4\tau}} {4\sin^2\frac{p_3}{2}+4\sin^2\frac{p_4}{2}}~.
\label{pprop}
\ee

\noi Perturbatively the gluon is massless, and for $\vec{p}=0$ this correlator
is ill--defined.

A naive implementation of the idea of dynamical mass generation
suggests to generalize eq.~(\ref{pprop}) as

\be
\Gamma_{\mu}(\vec{p},\tau) \longrightarrow \frac{3g^2}{2V_4}\sum_{p_4}
\frac{e^{-ip_4\tau}}
{4\sin^2\frac{p_3}{2}+4\sin^2\frac{p_4}{2}+4\sinh^2\frac{m_g}{2}}~,
\label{dmg}
\ee

\noi $m_g~$ being a `gluon mass' (defined such that $~E = m_g~$ for
$~{\vec p} =0~$). 
It is usual practice to interprete all deviations from eq.~(\ref{pprop}) 
as an effect of nonzero $~m_g~$ shown in eq.~(\ref{dmg})
(see, e.g. \cite{mo87}). In this work we would like to discuss
an alternative to eq.'s ~(\ref{pprop}) and ~(\ref{dmg}) for $\vec{p}=0$.

There is a degeneracy of the $~x$--independent (zero--momentum)
solutions $~U^{cl}_{x\mu}~$ of the classical equations of motion due to
the toroidal structure of the periodic lattice \cite{toro}. 
An example of such a solution with zero action (toron) is $~U^{cl}_{x\mu} =
e^{i\phi_{\mu}T}~$, where $~T~$ is one of the generators of the gauge
group and $~\phi_{\mu}~$ are four numbers. The perturbative expansion
deals with fluctuations about the classical solutions (e.g. torons),
and the shift $~A^a_{x\mu} \to \phi_{\mu} + A^a_{x\mu}~$ (for small
fields) produces the appearance  of a non--negative constant term
$~\sim \langle\phi_{\mu}^2\rangle~$ in the correlator 
$~\Gamma_{\mu}(\tau)$.

To get our alternative description to eq.'s ~(\ref{pprop}) and ~(\ref{dmg})
we split off the zero--momentum parts

\be
C_{\mu}=\frac{1}{V_4}\sum_{x}\cao_{\mu}(x)
\label{C}
\ee
of the operators $~\cao_{\mu}(x)~$
so that they decompose as

\be
\tao_{\mu}(\vec{0},\tau)= C_{\mu} + \delta\tao_{\mu}(\tau) \quad .
\label{del}
\ee

\noi Then eq.~(\ref{prop}) for $~\vp =0~$ gets the form 
\ba
\Gamma_{\mu}(\tau) &=& \mbox{Tr} \Bigl\langle C_{\mu}^2\Bigr\rangle
+ R_{\mu}(\tau)~;
\label{dprop}
\\
\nonumber \\
R_{\mu}(\tau) &=& \frac{1}{L_4}
\sum_t \mbox{Tr} \langle\delta\tao_{\mu}(t+\tau)
\delta\tao_{\mu}(t)\rangle \quad .
\label{rprop}
\ea

\noi Evidently,

\be
\frac{1}{L_4}\sum_{\tau} \Gamma_{\mu}(\tau)=\mbox{Tr}
\Bigl\langle C_{\mu}^2\Bigr\rangle~. 
\label{tprop}
\ee

\noi The first term on the r.h.s.~of (\ref{dprop}), being defined in terms
of the original fields, can be readily determined in the Monte Carlo
simulations. 
The term $R_{\mu}(\tau)$ in (\ref{dprop}) describes
correlations between fields $\delta\tao_{\mu}(\tau)$ with
zero-momentum parts subtracted. Making the change of variables
$~A_{x\mu}\to \phi_{\mu} + A_{x\mu}~$ 
and using a collective-coordinate method\footnote{which works here, of course,
only in lowest approximation}
one obtains in gaussian approximation
\be
\Gamma_{\mu}(\tau) = \mbox{Tr}\Bigl\langle C_{\mu}^2\Bigr\rangle
+ \frac{3g^2}{2V_4}\sum_{p_4\ne 0}
\frac{e^{-ip_4\tau}} {4\sin^2\frac{p_4}{2}}~,
\label{wprop}
\ee

\noi which will be our reference form in the numerical study of the
propagators $~\Gamma_{\mu}(\tau)$.

\section{Broken $~Z(2)~$ symmetry} \setcounter{equation}{0}

To monitor the breaking of the $~Z(2)~$ symmetry we have measured 
Polyakov loops
\be
P_{\mu}=\frac{L_{\mu}}{V_4}\sum_{x\ne x_{\mu}} 
\mbox{Tr}\prod_{x_{\mu}}U_{\mu x}~.
\ee
\noi Because of the two
possibilities for each direction (i.e., $P_{\mu}>0$ or $P_{\mu}<0$)
there are $2^4=16$  states : $(++++)$, $(+++-)$, $\ldots~$, $(----)$.
Fig.~1 shows the time history of the $P_{\mu}$, $\Gamma_T$ and
$\Gamma_4$ obtained on the $4^3\times 8$ lattice.  It is seen that the
$\Gamma_{\mu}$ are, in fact, strongly affected by the indicated states
and, in particular, that also the space-like directions are important
within this respect. The close relation between $P_{\mu}$ and $\Gamma_4$
is also obvious.

On larger lattices the same features are observed (though getting accurate
data for illustrations by time histories gets less easy). The phenomenon
occurs for all $\beta$ in the deconfinement region. The transition rate between
the states is smaller on larger lattices and for larger $\beta$.
It depends, of course, on the algorithm which is used. For example, with heat
bath updates there are considerably more transitions than with Metropolis.

 From the numerical analysis we find in more detail that the observables,
depending on their symmetry properties, take different values in the states.
Huge differences as well as smaller differences of the values taken by the
observables in different states are observed. This is seen in Fig.~2 for the
example of $\Gamma_T$ at finite $\vec{p}$ on the $16^3\times 32$ lattice
(for clarity of the figure the statistical errors are not shown; they range
from less than 2 \% for the $(++++)$--state to about 20 \% for states with the
largest magnitudes of the propagators).

It should be obvious that to obtain sensible results measurements of
observables are to be done separately for each state. In particular,
for comparison with usual perturbation theory only the $(++++)$--state is
appropriate. To improve statistics, instead of only selecting the
$(++++)$--data
from the time history, the other configurations (solely) for the purpose of
determining observables may be transformed to $(++++)$. We have checked that
this leads (faster) to the same results.
The Z(2) transformation used, which for a particular direction leads from 
$-$ to $+$ , is simply a factor $-1$ applied to the field variables of
this direction (for the time direction this has already been used in
Ref.~\cite{kr95}).

\section{Zero momentum modes} \setcounter{equation}{0}

\hspace{0.35cm}
We now concentrate on the $(++++)$--state at large $\beta$, i.e.~$\beta=10$
and $\beta=99$. To test the accuracy of the lowest order calculations we have
compared numerical results for $p_3\ne 0$ with (\ref{pprop}). The deviations
are generally smaller than about 10 \% and become smaller for larger $\beta$
and for larger $p_3$. They are certainly very small as compared to the
effects to be discussed below. Thus it appears that for the present study
we are sufficiently deep in the perturbative region.

 For $\vec{p}=\vec{0}$ the comparison with (\ref{wprop}) requires the
knowledge of $\langle C_{\mu}^2\rangle$ which we also determine in the
Monte Carlo simulations. In addition, we obtain important properties of
$C_{\mu}$ by comparing the time histories of $C_{\mu}$,
$\delta\tao_{\mu}(\tau)$ and of
\be
X_{\mu}(\rho)=
\mbox{Re }\frac{1}{V_4}\sum_{x}e^{i\vec{p}\cdot\vec{x}}\cao_{\mu}(x)
\ee
with $\vec{p}=(0,0,\frac{2\pi}{L_3}\rho$) and $\rho\ne 0$, i.e.~of an
example of a nonzero-momentum part of the fields.

Fig.~3 shows time histories of components $C_{\mu}^a$, $\delta \tao_{\mu}^a$
and $X_{\mu}^a$ (where $C_{\mu} =\sum_a (\sigma^a/2)C_{\mu}^a$ etc.). For
$\delta\tao_{\mu}^a$, $X_{\mu}^a(1)$ and $X_{\mu}^a(2)$ uniform Monte Carlo
noise is seen. Its magnitude is larger if the three-momentum involved in
the particular quantity gets smaller. The behavior of $\delta\tao_{\mu}^a$,
in Fig.~3 presented for $\tau=0$, is the same for all $\tau$.
For $C_{\mu}^a$ in addition to noise with magnitude comparable to that of
$\delta \tao_{\mu}^a$, surprisingly large variations are observed. These
variations exhibit different patterns for different $\mu$, while for different
$a$ we find the same pattern.

A very interesting aspect of these results is that in
the decomposition (\ref{del}) of the fields $\tao_{\mu}$ the parts
$C_{\mu}^a$ and $\delta \tao_{\mu}^a$ obviously behave quite differently.
The observed large variations of $C_{\mu}^a$ appear to be a characteristic
consequence of zero-momentum modes. They seem to be related to changes between
different constant classical solutions as one may consider in the
weak-coupling limit.

Because of the considerable size of the large variations of $C_{\mu}$ it is
to be expected that also $\langle C_{\mu}^2\rangle$ gets large as compared
to other quantities. This will be seen to be indeed the case below. A further
consequence of these variations is that error analyses of
observables involving $C_{\mu}$ give unusually large errors. This, in
particular holds for the propagator (\ref{dprop}). In order to show that
the unusually large errors are entirely due to the part
$\mbox{Tr} \langle C_{\mu}^2\rangle$ we also have measured $R_{\mu}(\tau)$
separately in the simulations. An example of the respective results is
shown in Fig.~4. It is seen that while $\Gamma_T(\vec{0}, \tau)$ exhibits 
very large errors, for $R_T(\tau)$
one indeed obtains errors of usual size.

 From Fig.~4 it is obvious that our numerical results for the transverse
propagator with $\vec{p}=\vec{0}$ agree reasonably well with that of
our lowest-order calculation (\ref{wprop}). The average over all times of
the propagator is precisely $\mbox{Tr} \langle C_{\mu}^2\rangle$ as predicted
by (\ref{tprop}). Analogous observations as on the $8^3\times 16$ lattice for
$\beta=10$ have been made on the $4^3\times 8$ lattice for $\beta=10$ and on
the $16^3\times 32$ lattice for $\beta=10$ and $\beta=99$. Thus our numerical
results confirm the description we have given.

To quantify the importance of zero-momentum modes we compare
$\mbox{Tr}\langle C_{\mu}^2\rangle$ with $R_{\mu}(0)$ (Table 1) and also with
$\Gamma_{\mu}(\vec{p},0)$ with $\vec{p} \ne \vec{0}$ (Table 2). To allow for
the comparison of different lattice sizes and different $\beta$ the values
in Tables 1 and 2 are multiplied by $V_3/g^2$ (which obviously cancels the
extra factor implicit in our definitions as compared to usual continuum
expressions). While all the other quantities then are roughly of the same
order of magnitude, the values of $\mbox{Tr} \langle C_{\mu}^2\rangle$
turn out to be much larger. From Table 1 it is seen that for increasing
$\beta$ this feature gets even more pronounced. The same holds for increasing
lattice size. Not only $\zeta$ gets larger but there is also an increase
of the ratios $\zeta/\gamma$. 

In literature investigations of the gluon mass have been based on
determinations of effective masses $~m(\tau )~$ from
\be
\frac{\cosh (m(\tau )(\tau +1-\frac{L_4}{2}))}
{\cosh (m(\tau )(\tau -\frac{L_4}{2}))}
= \frac{\Gamma_T(\vec{0},\tau +1)}{\Gamma_T(\vec{0},\tau )}~.
              \label{m}
\ee

\noi  For the present data the respective results are depicted in Figure~5. 
Because $~m(\tau)L_{\mu}~$ shows little dependence on lattice size,
these results may also be considered from the point of view of 
finite temperatures where screening masses are determined \cite{hkr95,hkr97}.
We have shown that instead of referring to effective masses the data
can perfectly be described by eq.~(\ref{wprop}) relying on zero--momentum
modes.

\section{Discussion and conclusions}

To summarize, the dependence of the gauge--variant (zero momentum)
propagators $~\Gamma_{\mu}(\tau)~$ on zero--momentum modes
can be very strong. The mechanism of this influence can be viewed as follows.
The chosen gauge fixing does not prohibit the fluctuations about
some nonzero constant values $~\phi^a_{\mu} \ne 0~$, and the
shift $~A^a_{x\mu}\to\phi^a_{\mu}+A^a_{x\mu}~$ (for small fields)
produces the appearence of a non--negative constant term in
the correlator $~\Gamma_{\mu}(\tau) \to \mbox{Const} + R_{\mu}(\tau)~$.
In this paper we demonstrated the occurrence of a large constant term
explicitly in the weak coupling (perturbative) limit.

Considering time histories in our simulations we have found that the
zero--momentum term shows unusually large fluctuations. These
fluctuations appear to be a characteristic signal for zero--momentum modes.

The question of the role of the zero--momentum modes at smaller 
values of $~\beta~$ (in the physical region) needs further study.
Indeed, one expects dynamical mass generation in the 
physical (nonperturbative) regime. However, zero--momentum modes
which appear in the periodic volume may (at least, partially)
mimic it, as it happens in the pure gauge $~U(1)~$ theory.
The latter can be seen from the data in \cite{mo2} as has been discussed
in \cite{m97}.
To disentangle these two effects, i.e. dynamically generated 
mass and zero--momentum mode contribution, a comparative 
study with different boundary conditions could be of use.

On the finite lattice at large $~\beta~$ values one is beyond the 
finite--volume transitions, and thus  the propagators become 
sensitive to the broken $~Z(2)~$ symmetry states of the 
deconfinement region usually monitored by Polyakov loops.
We find here that the states signalled by the space--like 
Polyakov loops are important, too.


\section*{Acknowledgments}

\hspace{3mm}
One of us (W.K.)
wishes to thank M.~M\"uller-Preussker and his group for their kind
hospitality.
This research was supported in part under DFG grants Ke 250/13-1 and
Mu 932/1-4 and the grant INTAS--96--370.

\newpage
\renewcommand{\baselinestretch}{1.3}

\newpage
\renewcommand{\baselinestretch}{1.5}
\small\normalsize

\begin{center}

{\bf Table 1}

\vspace*{3mm}
$\zeta=(V_3/g^2) \mbox{Tr} \langle (C_T)^2\rangle$ and
$\gamma = (V_3/g^2)((R_T(0)- R_T(L_4/2))$
\vspace*{3mm}

\begin{tabular}{|c|c|c|c|}
\hline
 lattice      & $\beta$ & $\zeta$ &  $\gamma$  \\
\hline
$4^3\times 8$ &  10  &  6.30 (16)  &  1.610 (5)   \\
$8^3\times 16$ &  10 &  15.7 (7) &  3.229 (17)   \\
$16^3\times 32$ &  10 &  57 (10) &  6.03 (22)     \\
$16^3\times 32$ &  99 & 375 (86) &  4.98 (19)     \\
\hline
\end{tabular}
\vspace*{2.5cm}

{\bf Table 2}

\vspace*{3mm}
$(V_3/g^2) \Gamma_T(\vec{p},0)$ with $\vec{p}=(0,0,\frac{2\pi}{L_3}\rho)$
\vspace*{3mm}

\begin{tabular}{|c|c|c|c|c|}
\hline
 lattice      & $\beta$ & $\rho=1$ & $\rho=2$ & $\rho=4$ \\
\hline
$4^3\times 8$ &  10    & 0.460 (2)&  0.275 (1)     &            \\
$8^3\times 16$ &  10   & 0.992 (4)&  0.457 (1)     &  0.277 (1)   \\
$16^3\times 32$ &  10  & 2.100 (20)&  0.981 (4)    &  0.457 (1)    \\
$16^3\times 32$ &  99  & 2.002 (23)&  0.926 (5)    &  0.436 (2)     \\
\hline
\end{tabular}

\end{center}

\newpage
\renewcommand{\baselinestretch}{1.7}
\small\normalsize

\section*{Figure captions}

\begin{tabular}{rl}

Fig.~1. & Time history of Polyakov loops and propagators (at $\tau=0$)\\
	& for $\beta=10$ and $4^3\times 8$ lattice.\\

Fig.~2. & Transverse propagators in various states for
          $\vec{p}=(0,0,\frac{2\pi}{L_3})$ ,\\
	& $\beta=10$ and $16^3\times 32$ lattice
	  (curve is (\ref{pprop})).\\

Fig.~3. & Time histories of $C_{\mu}^a$, $\delta \tao_{\mu}^a$,
	  $X_{\mu}^a(1)$ and $X_{\mu}^a(2)$ \\
	& for $\beta=10$ and $16^3\times 32$ lattice.\\

Fig.~4. & $\Gamma_T(\vec{0},\tau)$ and $R_T(\tau)$
          compared with (\ref{wprop}) (curves) and
	  $\mbox{Tr} \langle C^2_{\mu}\rangle$\\
        & (constant line in upper Figure)
	  for $\beta=10$ and $8^3\times 16$ lattice.\\
        & Errors of $R_T$ are smaller than symbols. \\

Fig.~5. & Effective masses from eq.~(\ref{m}) for $\vec{p}=\vec{0}$, $\beta=10$
	  and  \\
        & various lattice sizes (curve based on (\ref{wprop})).

\end{tabular}

\newpage
\renewcommand{\baselinestretch}{1.3}
\small\normalsize

\begin{figure}[tb]
\centering
\epsfig{file=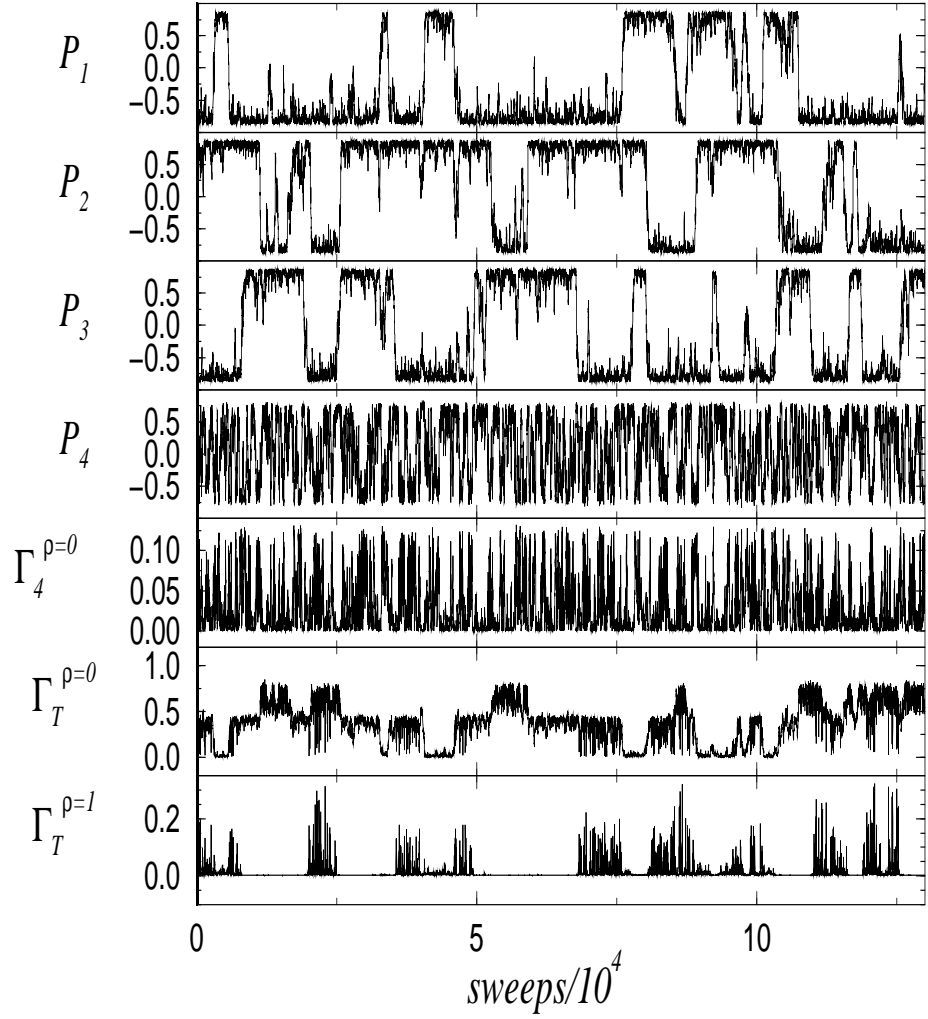,width=13cm,height=16cm}
\caption{
Time history of Polyakov loops and propagators at $\tau=0$
for $\beta=10$ and $4^3\times 8$ lattice.
}
\end{figure}

\begin{figure}[tb]
\centering
\epsfig{file=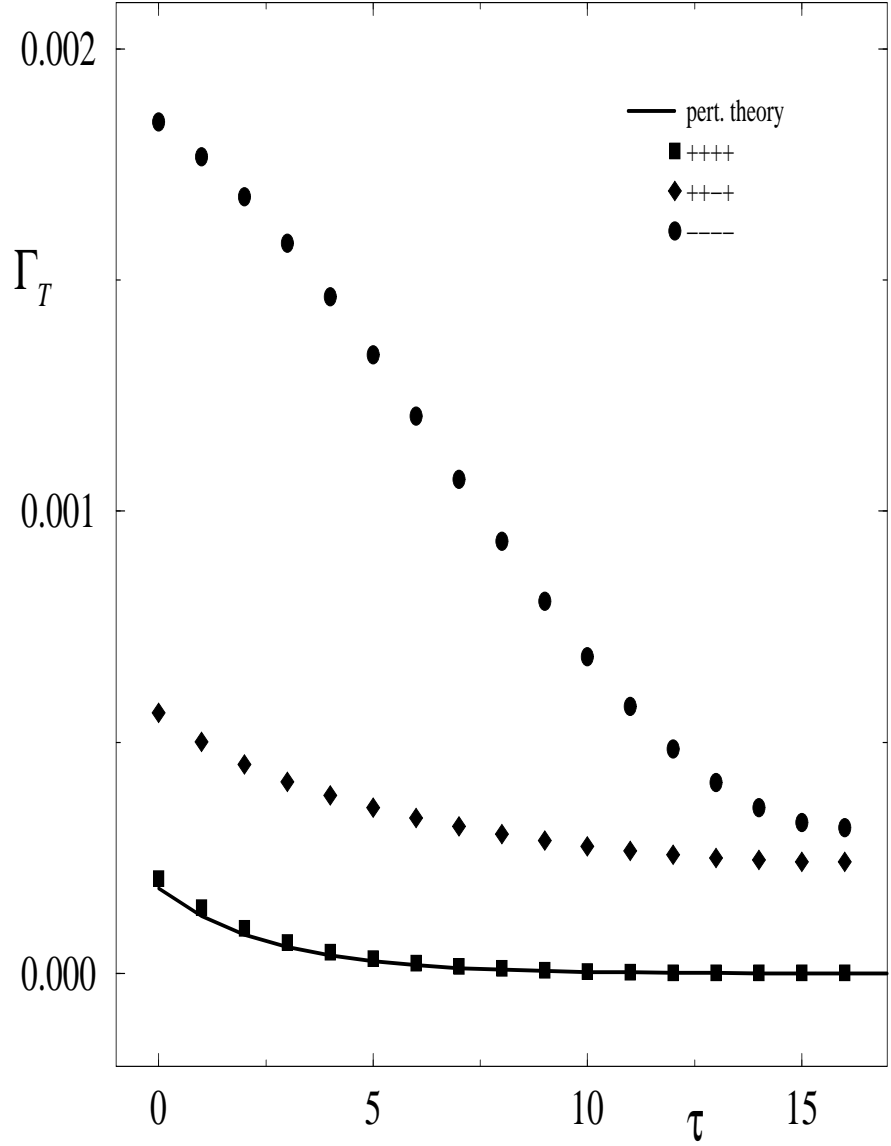,width=13cm,height=16cm}
\caption{
Transverse propagators in various states for
$\vec{p}=(0,0,\frac{2\pi}{L_3})$ ,
$\beta=10$ and $16^3\times 32$ lattice
(curve is (\ref{pprop})).
}
\end{figure}

\begin{figure}[tb]
\centering
\epsfig{file=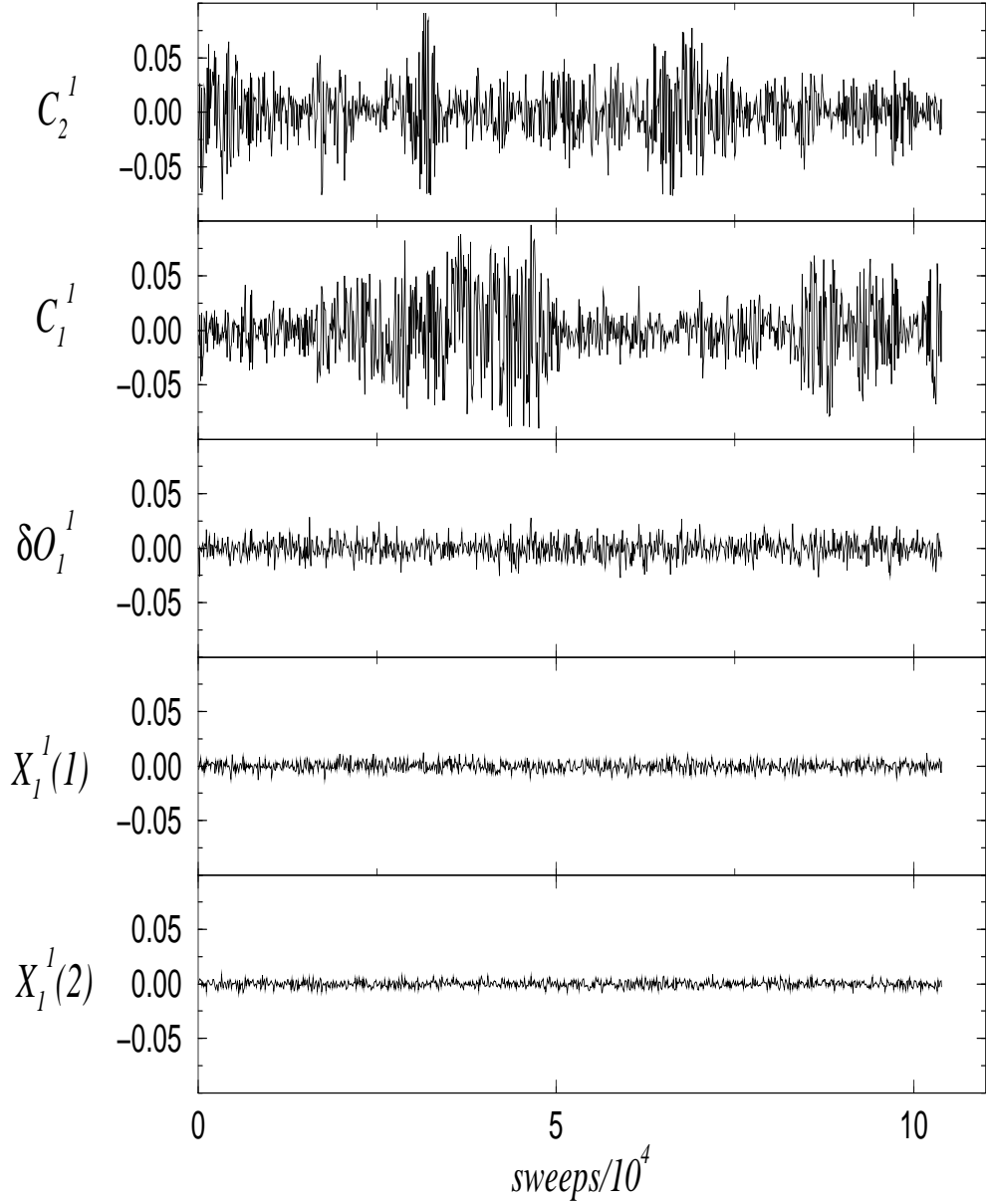,width=13cm,height=16cm}
\caption{
Time histories of $C_{\mu}^a$, $\delta \tao_{\mu}^a$,
$X_{\mu}^a(1)$ and $X_{\mu}^a(2)$
for $\beta=10$ and $16^3\times 32$ lattice.
}
\end{figure}

\begin{figure}[tb]
\centering
\epsfig{file=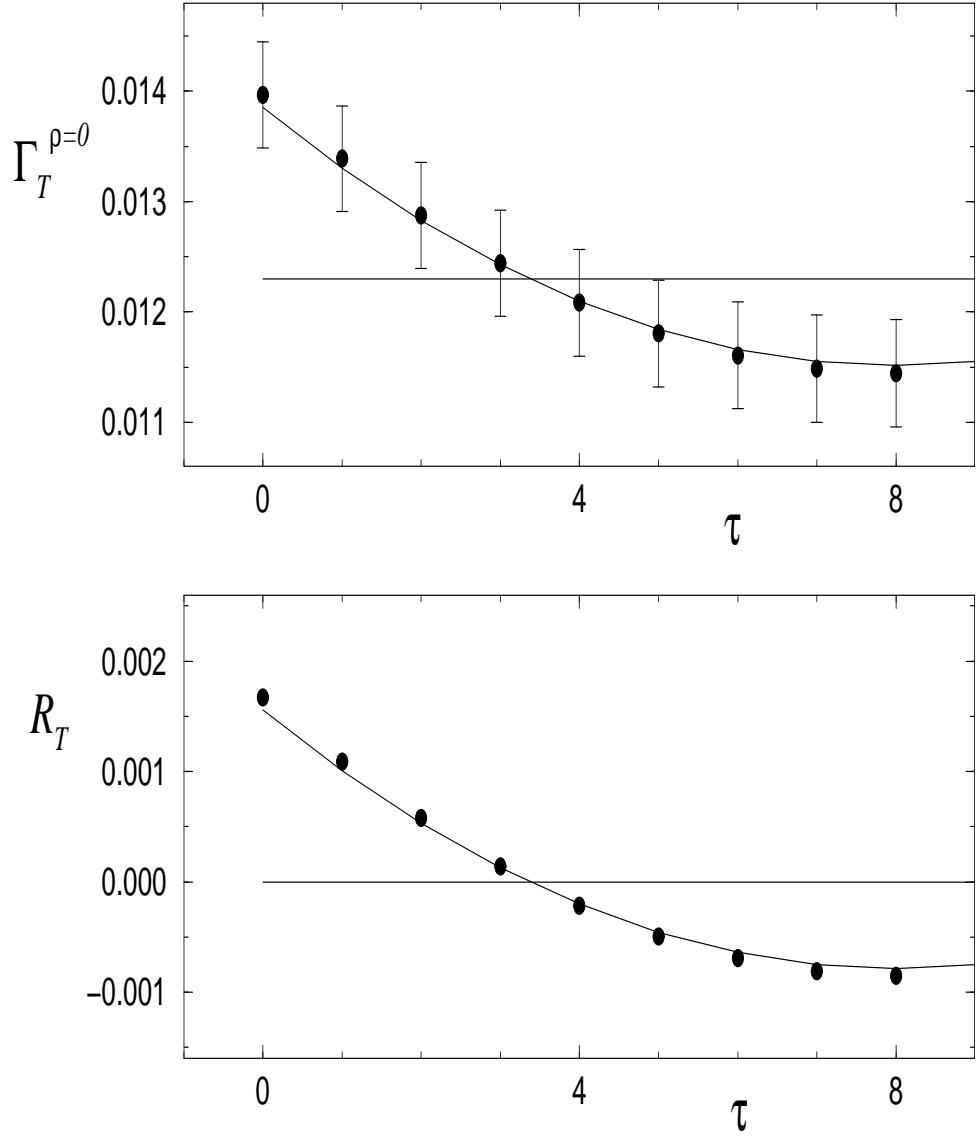,width=13cm,height=16cm}
\caption{
$\Gamma_T(\vec{0},\tau)$ and $R_T(\tau)$
compared with (\ref{wprop}) (curves) and
$\mbox{Tr} \langle C^2_{\mu}\rangle$
(constant line in upper Figure)
for $\beta=10$ and $8^3\times 16$ lattice.
}
\end{figure}

\begin{figure}[tb]
\centering
\epsfig{file=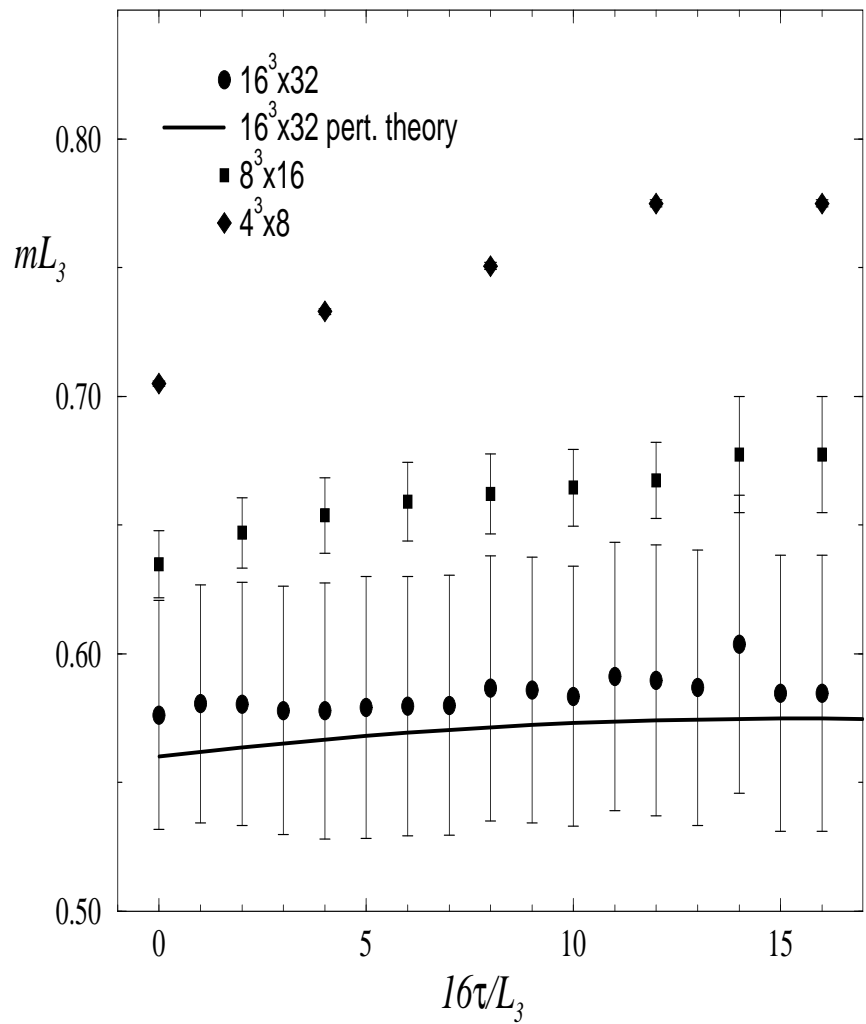,width=13cm,height=16cm}
\caption{
Effective masses from eq.~(\ref{m}) for $\vec{p}=\vec{0}$, $\beta=10$
and various lattice sizes (curve based on (\ref{wprop})).
}
\end{figure}

\end{document}